\documentclass[
]{ceurart}

\usepackage{listings}
\usepackage[most]{tcolorbox}
\usepackage{lipsum}
\usepackage{caption}
\usepackage{amsmath}

\usepackage{glossaries}
\setacronymstyle{long-short}
\newacronym{nlp}{NLP}{natural language processing}
\newacronym{sp}{SP}{sound processing}
\newacronym{nn}{NNs}{neural networks}
\newacronym{dl}{DL}{deep learning}
\newacronym{dnns}{DNNs}{deep neural networks}
\newacronym{ml}{ML}{machine learning}
\newacronym{tl}{TL}{transfer learning}
\newacronym{ws}{WS}{weak supervision}
\newacronym{al}{AL}{active learning}
\newacronym{dal}{DAL}{deep active learning}
\newacronym{mlp}{MLP}{multi-layer perceptron}
\newacronym{dnn}{DNN}{deep neural network}
\newacronym{sota}{SOTA}{state-of-the-art}
\newacronym{cv}{CV}{computer vision}
\newacronym{plm}{PLMs}{pre-trained language models}
\newacronym{xc}{XC}{Xeno-Canto}
\newacronym{pam}{PAM}{passive acoustic monitoring}
\newacronym{sed}{SED}{sound event detection}
\newacronym{cnn}{CNN}{convolutional neural network}
\newacronym{rnn}{RNN}{recurrent neural network}
\newacronym{ast}{AST}{audio spectrogram transformer}
\newacronym{w2v2}{W2V2}{Wav2Vec 2.0}

\newcommand{\extractorparams}{\boldsymbol{\omega}}

\begin{document}

\copyrightyear{2024}
\copyrightclause{Copyright for this paper by its authors.
  Use permitted under Creative Commons License Attribution 4.0
  International (CC BY 4.0).}

\conference{IAL@ECML-PKDD'24:
  8\textsuperscript{th} Intl. Worksh. \& Tutorial on Interactive Adaptive Learning,
  Sep. 9\textsuperscript{th}, 2024, Vilnius, Lithuania}

\title{Towards Deep Active Learning in Avian Bioacoustics}
% \title{A Pilot Study on Deep Active Learning in Avian Bioacoustics for Passive Acoustic Monitoring}

% \tnotetext[1]{You can use this document as the template for preparing your
%   publication. We recommend using the latest version of the ceurart style.}

\author[1]{Lukas Rauch}[%
email=lukas.rauch@uni-kassel.de,
]
%\cormark[1]
\address[1]{IES, University of Kassel, Kassel, Germany}
\address[2]{IEE, Fraunhofer Insitute, Kassel, Germany}

\author[1]{Denis Huseljic}[%
]
\author[1]{Moritz Wirth}[%
]
\author[1]{Jens Decke}[%
]
\author[1]{Bernhard Sick}[%
]
\author[1]{Christoph Scholz}[%
]
% \author[1]{Anonymous Author}[%
% email= anonymous@anonymous.de
% ]
% %\cormark[1]
% %\fnmark[1]
% \address[1]{Anonymous Institute}
% \address[2]{Anonymous Institute}

% \author[1]{Denis Huseljic}[%
% ]
% \author[1]{Moritz Wirth}[%
% ]
% \author[1]{Jens Decke}[%
% ]
% \author[1]{Bernhard Sick}[%
% ]
% \author[1]{Christoph Scholz}[%
% ]

% Anonymous Authors
%\cortext[1]{Corresponding author.}
%\fntext[1]{These authors contributed equally.}

\begin{abstract}
Passive acoustic monitoring (PAM) in avian bioacoustics enables cost-effective and extensive data collection with minimal disruption to natural habitats. Despite advancements in computational avian bioacoustics, deep learning models continue to encounter challenges in adapting to diverse environments in practical PAM scenarios. This is primarily due to the scarcity of annotations, which requires labor-intensive efforts from human experts. Active learning (AL) reduces annotation cost and speed ups adaption to diverse scenarios by querying the most informative instances for labeling. This paper outlines a deep AL approach, introduces key challenges, and conducts a small-scale pilot study.
\end{abstract}

\begin{keywords}
  Deep Active Learning \sep
  Avian Bioacoustics \sep
  Passive Acoustic Monitoring
\end{keywords}

\maketitle
\section{Introduction}
Avian diversity is a key indicator of environmental health. \Gls*{pam} in avian bioacoustics leverages mobile autonomous recording units (ARUs) to gather large volumes of soundscape recordings with minimal disruption to avian habitats. While this method is cost-effective and minimally invasive, the analysis of these recordings is labor-intensive and requires expert annotation. Recent advancements in \gls*{dl} primarily process these passive recordings by classifying bird vocalizations. Particularly, feature embeddings from large bird sound classification models (e.g., Google's Perch~\cite{hamer2023birb} or BirdNET~\cite{kahl2021birdnet}) have effectively enabled few-shot learning in scenarios with limited training data~\cite{ghani2023}. These \gls*{sota} models are trained using supervised learning on nearly 10,000 bird species from multi-class focal recordings that isolate individual bird sounds. However, practical \gls*{pam} scenarios involve processing diverse multi-label soundscapes with overlapping sounds and varying background noise. Proper feature embeddings for edge deployment necessitate fine-tuning, which relies on labeled training data that is both time-consuming and costly to obtain for soundscapes.

Deep \gls*{al} addresses this challenge by actively querying the most informative instances to maximize performance gains~\cite{decke2023dado}. However, research on deep \gls*{al} in avian bioacoustics is still limited, and the problem needs to be contextualized with comparable datasets~\cite{rauch2023b2v}. Additionally, the domain presents unique practical challenges, including adapting models from focals to soundscapes (i.e., multi-class to multi-label) in imbalanced and highly diverse scenarios~\cite{rauch2024birdset}. Consequently, we introduce the problem of deep \gls*{al} in avian bioacoustics and propose an efficient fine-tuning approach for model deployment. Our contributions are:

\begin{tcolorbox}[title=Contributions, arc=0pt, boxrule=0.5pt, left=2pt, top=2pt, bottom=2pt]
\begin{enumerate}
    \item{We introduce deep active learning (AL) to avian bioacoustics, highlighting challenges and proposing a practical framework.}
    % \item{We propose and work out different approaches and scenarios for deep AL in this domain.}
    \item{We conduct an initial feasibility study based on the dataset collection \texttt{Birdset}~\cite{rauch2024birdset}, showcasing the benefits of deep AL. Additionally, we release the dataset and code.}
\end{enumerate}
\end{tcolorbox}

\section{Related Work}
\Gls*{dl} has enhanced bird species recognition from vocalizations in the context of biodiversity monitoring. Current \gls*{sota} approaches BirdNET~\cite{kahl2021birdnet}, Google's Perch~\cite{denton2021separation,hamer2023birb}, and \texttt{BirdSet}~\cite{rauch2024birdset} have set benchmarks in bird sound classification. While initial studies focused on model performance on focal recordings, research is increasingly shifting towards practical \gls*{pam} scenarios~\cite{rauch2024birdset}. In such environments, ARUs are proving effective for edge deployment for continuous soundscape analysis~\cite{hoechst22}. Research indicates that pre-trained models facilitate few-shot and transfer learning in data-scarce environments by providing valuable feature embeddings for rapid prototyping and efficient inference~\cite{ghani2023}. While deep \gls*{al} is suited for quick model adaptation, its application in avian bioacoustics is still emerging. \citet{bellafkir2023} have integrated \gls*{al} into edge-based systems for bird species identification, employing reliability scores and ensemble predictions to refine misclassifications through human feedback. This approach highlights the necessity for research into the application of deep \gls*{al} and multi-label classification in avian bioacoustics. However, comparing these results is challenging because they utilize test datasets that are not publicly available and employ custom \gls*{al} strategies~\cite{bellafkir2023}.

\section{Active Learning in Bird Sound Classification}
\label{sec:problem}

% A $D$-dimensional instance is represented by a feature vector $\mathbf{x}\in\mathcal{X}$ with the feature space $\mathcal{X} = \mathbb{R}^D$. 
{\noindent \textbf{Motivation}.} In \gls*{pam}, a feature vector $\mathbf{x} \in \mathcal{X}$ represents a $D$-dimensional instance, originating from either a focal recording where $\mathcal{X} = \mathcal{F}$, or a soundscape recording with $\mathcal{X} = \mathcal{S}$. \textbf{Focal recordings} are extensively available on the citizen-science platform \gls*{xc}~\cite{vellinga2015} with a global collection of over 800,000 recordings, making them particularly suitable for model training. Large-scale bird sound classification models (e.g., BirdNET\cite{kahl2021birdnet}) are primarily trained on focals. These multi-class recordings feature isolated bird vocalizations where each instance $\mathbf{x}$ is associated with a class label $y \in \mathcal{Y}$, where $\mathcal{Y}=\{1,...,C\}$. The focal data distribution is denoted as $p_{\texttt{Focal}}( \mathbf{x},y)$. However, annotations from \gls*{xc} often come with weak labels, lacking precise vocalization timestamps. As noted by \citet{vanmerrienbour2024}, evaluating on focals does not adequately reflect a model's generalization performance in real-world \gls*{pam} scenarios, rendering them unsuitable for assessing deployment capabilities.  
\textbf{Soundscape recordings} are passively recorded in specific regions, capturing the entire acoustic environment for \gls*{pam} projects using static ARUs over extended periods. For instance, the High Sierra Nevada (HSN)~\cite{kahl2021birdnet} dataset includes long-duration soundscapes with precise labels and timestamps from multiple recording sites. Soundscapes are treated as multi-label tasks and are valuable for assessing model deployment in real-world \gls*{pam}. Each instance $\mathbf{x}$ is associated with multiple class labels $y \in \mathcal{Y}$, represented by a one-hot encoded multi-label vector $\mathbf{y} = [y_1, \ldots, y_C] \in {[0, 1]}^C$. An instance can contain no bird sounds, represented by a zero-vector $\mathbf{y} = \mathbf{0} \in \mathbb{R}^C$. Soundscapes' limited scale and the extensive annotation effort make them less suitable for large-scale model training. We denote the soundscape data distribution as $p_{\texttt{Scape}}(\mathbf{x},\mathbf{y})$. The disparity in data distributions, $p_{\texttt{Scape}}(\mathbf{x},\mathbf{y}) \neq p_{\texttt{Focal}}(\mathbf{x},y)$, leads to a distribution shift that impacts the performance of \gls*{sota} bioacoustic models trained on focals when deployed in \gls*{pam}. Additionally, highly diverse deployment conditions in \gls*{pam} projects - such as background noise, recording devices, and their locations - also lead to domain differences within and between soundscape recordings. These variations further highlight the need for compact models that can quickly and easily adapt to changing environments. Thus, we argue that using labeled soundscapes in novel deployment scenarios for fine-tuning the model is vital. Therefore, we propose deep \gls*{al} to enable fast model adaption to various \gls*{pam} scenarios.

\begin{figure*}[!h]
\centering
\includegraphics[width=0.95\columnwidth]{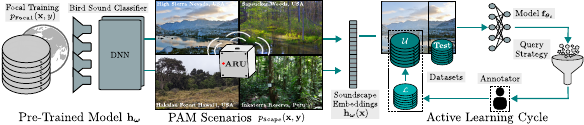}
\center\caption{Proposed deep AL cycle in avian bioacoustics with exemplary tasks from \texttt{BirdSet}\cite{rauch2024birdset}.}
\label{fig:GA}
\vspace{-0.45cm}
\end{figure*}

{\noindent\textbf{{Our approach.}}}
 Our approach is detailed in \autoref{fig:GA}. We leverage the \texttt{BirdSet} dataset collection \cite{rauch2024birdset} to ensure comparability. 
% The soundscape recordings are separated into 5-second segments that serve as an instance.
We consider a multi-label classification problem, where we equip a model with a pre-trained feature extractor $\mathbf{h}_{\extractorparams}: \mathcal{X} \to \mathbb{R}^D$ with parameters $\extractorparams$ that maps the inputs $\mathbf{x}$ to feature embeddings $\mathbf{h}_{\extractorparams}(\mathbf{x})$. Additionally, we utilize a classification head $\mathbf{f}_{\boldsymbol{\theta}_{t}}: \mathbb{R}^D \to \mathbb{R}^C$ with parameters $\boldsymbol{\theta}_t$ at cycle iteration $t$ that maps the feature embeddings $\mathbf{h}_{\extractorparams}(\mathbf{x})$ to class probabilities via the sigmoid function. The resulting class probabilities are denoted by $\hat{\mathbf{p}} = \sigma(\mathbf{f}_{\boldsymbol{\theta}_{t}}(\mathbf{h}_{\extractorparams}(\mathbf{x}))$, where $\hat{\mathbf{p}} \in \mathbb{R}^C$ represents the probabilities for each class in a binary classification problem. We introduce a pool-based AL setting with an unlabeled pool ${\mathcal{U}(t) \subseteq \mathcal{S}}$ and a labeled pool data set ${\mathcal{L}(t) \subseteq \mathcal{S} \times \mathcal{Y}}$. The pool consists of soundscapes from \gls*{pam} projects, allowing the model to adapt to the unique acoustic features of new sites and improve performance across various scenarios. During each cycle iteration $t$, the query strategy compiles the most informative instances into a batch ${\mathcal{B}(t) \subset \mathcal{U}(t)}$ of size $b$. We represent an annotated batch as $\mathcal{B}^*(t) \in \mathcal{S} \times \mathcal{Y}$. We update the unlabeled pool $\mathcal{U}(t{+}1) = \mathcal{U}(t) \setminus \mathcal{B}(t)$ and the labeled pool $\mathcal{L}(t{+}1) = \mathcal{L}(t) \cup \mathcal{B}^*(t)$ by adding the annotated batch. At each iteration $t$, the model $\boldsymbol{\theta}_{t}$ is retrained using the binary cross entropy loss $L_{BCE}(\mathbf{x,y)}$, resulting in the updated model parameters $\boldsymbol{\theta}_{t+1}$. The process continues until a budget~$B$ is exhausted. 

\section{Experiments}
\noindent{\textbf{Setup.}} We employ Google's Perch as the pre-trained feature extractor with a feature dimensionality of $D=1280$, following \citet{ghani2023}. Each iteration of the \gls*{al} cycle involves initializing and training the last DNN layer for 200 epochs using the Rectified Adam optimizer~\cite{liu2019radam} (batch size: 128, learning rate: 0.05, weight decay: 0.0001) with a cosine annealing scheduler~\cite{huseljic2024fast}. The hyperparameters are empirically determined with convergence on random train samples as done in~\cite{huseljic2023role}. We utilize the HSN dataset~\cite{sierra_nevada_stefan_kahl_2022_7050014} from \texttt{BirdSet}~\cite{rauch2024birdset}, consisting of $5,280$ 5-second soundscape segments from the initial three days of recordings for our unlabeled pool. Thus, we simulate practical deployment scenario where we initially collect data from various recording sites that we want to quickly adapt the model to and reduce annotation effort. Subsequently, we utilize $6,720$ segments from the last two days for testing model performance. Initially, $10$ instances are selected randomly, followed by 50 iterations of $b{=}10$ acquisitions each, totaling a budget of ${B}{=}510$. We benchmark against \texttt{Random} acquisitions and use \texttt{Typiclust}~\cite{hacohen2022typiclust} and \texttt{Badge}\cite{ash2020BADE} as diversity-based and hybrid strategies, respectively. As an uncertainty-based strategy, we employ the mean \textsc{Entropy} of all binary predictions. The effectiveness of each strategy is assessed by analyzing the learning curves through a collection of threshold-free metrics~\cite{rauch2024birdset}: T1-accuracy, class-based mean average precision (cmAP), and area under the receiver operating characteristic curve (AUROC). The metrics are computed on the test dataset post-training in each cycle, with learning curve improvements averaged over ten repetitions for consistency.

\begin{figure*}[!h]
\centering
\includegraphics[width=0.9\columnwidth]{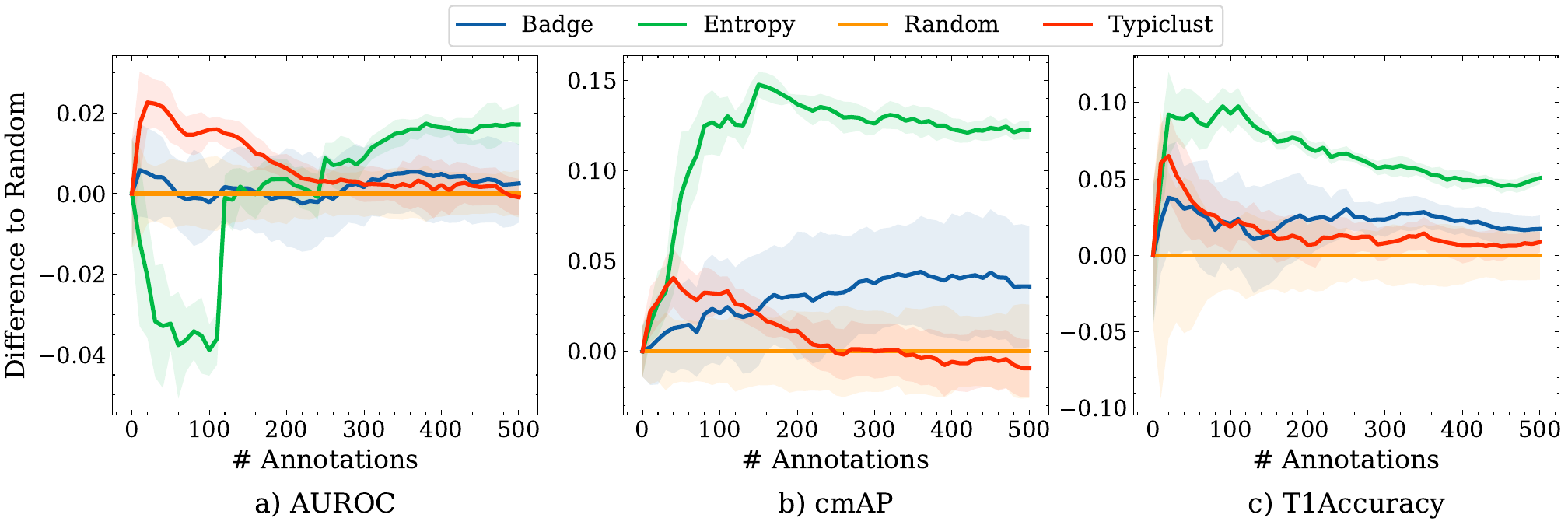}
\caption{Improvement curves of deep \gls*{al} selection strategies \texttt{Badge}, \texttt{Entropy}, and \texttt{Typiclust} over \texttt{Random} with the metric collection a)~AUROC, b)~cmAP and c)~T1-Acc. The results are averaged over ten randomly initialized repetitions to ensure consistency and the standard deviation is displayed.}
\label{fig:results}
\vspace{-0.6cm}
\end{figure*}

\noindent{\textbf{Results.}} 
We present the improvement curves for the metric collection in \autoref{fig:results}. The results demonstrate that no single strategy is universally superior across all metrics. However, nearly all metrics show enhanced performance compared to \texttt{Random}. Notably, \texttt{Typiclust} displays strong performance across all metrics at the start of the deep \gls*{al} cycle, supporting the findings of \cite{hacohen2022typiclust} that a diverse selection is beneficial at the cycle's onset. However, its effectiveness diminishes over time when diversity becomes less crucial. Conversely, except for the AUROC metric where \texttt{Entropy} initially performs poorly but strongly improves over time, \texttt{Entropy} outperforms in all iterations for cmAP and T1-Acc, showing a consistent improvement over \texttt{Random} of up to 15\%.

\section{Open Challenges and Limitations}

This pilot study explores the use of deep AL to tailor avian bioacoustic models for various deployment scenarios in PAM. Although the initial results are encouraging, they remain preliminary. Several key challenges, which are outlined below, need to be addressed to fully realize the potential of deep AL in this field.\vspace{0.3cm}

\noindent{\textbf{Pool creation.}} The limited availability of soundscape data, which is primarily used for model evaluation \cite{rauch2024birdset}, poses challenges in creating pool datasets for deep \gls*{al}. The process of generating a fine-tuning training pool can affect class balance and raises concerns about the composition methodology. Additionally, in scenarios where data are sourced from \gls*{pam} projects, the variability in recording sites is often not disclosed in publicly available datasets. This lack of information makes it challenging to create a diverse and representative training pool that takes recording locations into account. To effectively investigate deep AL, a transparent approach to dataset generation is essential.
\vspace{0.3cm}

\noindent{\textbf{Deployment in practice.}} Deploying deep \gls*{al} in real-world \gls*{pam} environments requires addressing several practical considerations. These include determining optimal batch sizes for data annotation and effectively allocating the total budget. The labor-intensive and costly process of labeling \gls*{pam} recordings, which requires human expertise \cite{stowell2021}, highlights the need for accurately estimating the expected annotation effort. Additionally, exploring various deployment settings and tasks can reveal the versatility and potential challenges of applying deep \gls*{al}, leading to more effective and scalable solutions for avian bioacoustics. For instance, tasks might involve not only classifying bird species but also identifying specific call densities \cite{navine2024}, which would require modifications to the model evaluation process. \vspace{0.3cm}

\noindent{\textbf{Evaluation.}} Traditional metrics such as AUROC, cmAP, and T1-Acc offer a general overview of model performance but may be inadequate in practice-specific scenarios, such as ensuring a high recall of a specific species or identifying bird call density \cite{navine2024}. A more nuanced approach to evaluating deep \gls*{al} models involves customizing metrics to align with practical objectives, such as consistently identifying specific species. Enhancing evaluation methodologies to capture these specialized requirements is crucial for advancing the effectiveness of deep \gls*{al} in real-world PAM applications. \vspace{0.3cm}

\section{Conclusion}
In this work, we demonstrated the potential of deep active learning (AL) in computational avian bioacoustics. We showed how deep AL can be integrated into real-world passive acoustic monitoring by utilizing \texttt{BirdSet}, where a rapid model adaption through fine-tuning on soundscape recordings is advantageous for the identification of bird species. Our results indicate that employing selection strategies in deep AL enhances model performance and accelerates adaptation compared to random sampling. For future work, we aim to expand the implementation of deep AL in avian bioacoustics utilizing all datasets from the \texttt{BirdSet} dataset collection to provide more robust performance insights and explore additional query strategies \cite{huseljic2024fast,rauch2023GLAE}.
%Additionally, we plan to identify a robust evaluation protocol to utilize the publicly available soundscape recordings. 

%%
%% Define the bibliography file to be used
\bibliography{literatur.bib}

\begin{thebibliography}{20}
\expandafter\ifx\csname natexlab\endcsname\relax\def\natexlab#1{#1}\fi
\providecommand{\url}[1]{\texttt{#1}}
\providecommand{\href}[2]{#2}
\providecommand{\path}[1]{#1}
\providecommand{\DOIprefix}{doi:}
\providecommand{\ArXivprefix}{arXiv:}
\providecommand{\URLprefix}{URL: }
\providecommand{\Pubmedprefix}{pmid:}
\providecommand{\doi}[1]{\href{http://dx.doi.org/#1}{\path{#1}}}
\providecommand{\Pubmed}[1]{\href{pmid:#1}{\path{#1}}}
\providecommand{\bibinfo}[2]{#2}
\ifx\xfnm\relax \def\xfnm[#1]{\unskip,\space#1}\fi
%Type = Article
\bibitem[{Hamer et~al.(2023)Hamer, Triantafillou, {Van Merri{\"e}nboer}, Kahl, Klinck, Denton, and Dumoulin}]{hamer2023birb}
\bibinfo{author}{J.~Hamer}, \bibinfo{author}{E.~Triantafillou}, \bibinfo{author}{B.~{Van Merri{\"e}nboer}}, \bibinfo{author}{S.~Kahl}, \bibinfo{author}{H.~Klinck}, \bibinfo{author}{T.~Denton}, \bibinfo{author}{V.~Dumoulin},
\newblock \bibinfo{title}{{{BIRB}}: {{A Generalization Benchmark}} for {{Information Retrieval}} in {{Bioacoustics}}},
\newblock \bibinfo{journal}{CoRR}  (\bibinfo{year}{2023}). \URLprefix \url{https://doi.org/10.48550/arXiv.2312.07439}.
%Type = Article
\bibitem[{Kahl et~al.(2021)Kahl, Wood, Eibl, and Klinck}]{kahl2021birdnet}
\bibinfo{author}{S.~Kahl}, \bibinfo{author}{C.~M. Wood}, \bibinfo{author}{M.~Eibl}, \bibinfo{author}{H.~Klinck},
\newblock \bibinfo{title}{{{BirdNET}}: {{A}} deep learning solution for avian diversity monitoring},
\newblock \bibinfo{journal}{Ecological Informatics} \bibinfo{volume}{61} (\bibinfo{year}{2021}) \bibinfo{pages}{101236}. \URLprefix \url{https://doi.org/10.1016/j.ecoinf.2021.101236}.
%Type = Article
\bibitem[{Ghani et~al.(2023)Ghani, Denton, Kahl, and Klinck}]{ghani2023}
\bibinfo{author}{B.~Ghani}, \bibinfo{author}{T.~Denton}, \bibinfo{author}{S.~Kahl}, \bibinfo{author}{H.~Klinck},
\newblock \bibinfo{title}{Global birdsong embeddings enable superior transfer learning for bioacoustic classification},
\newblock \bibinfo{journal}{Scientific Reports} \bibinfo{volume}{13} (\bibinfo{year}{2023}). \URLprefix \url{http://dx.doi.org/10.1038/s41598-023-49989-z}. \DOIprefix\doi{10.1038/s41598-023-49989-z}.
%Type = Inproceedings
\bibitem[{Decke et~al.(2023)Decke, Gruhl, Rauch, and Sick}]{decke2023dado}
\bibinfo{author}{J.~Decke}, \bibinfo{author}{C.~Gruhl}, \bibinfo{author}{L.~Rauch}, \bibinfo{author}{B.~Sick},
\newblock \bibinfo{title}{{DADO} {-}{-} {L}ow-cost query strategies for deep active design optimization},
\newblock in: \bibinfo{booktitle}{2023 International Conference on Machine Learning and Applications (ICMLA)}, \bibinfo{organization}{IEEE}, \bibinfo{year}{2023}, pp. \bibinfo{pages}{1611--1618}.
%Type = Article
\bibitem[{Rauch et~al.(2023)Rauch, Schwinger, Wirth, Sick, Tomforde, and Scholz}]{rauch2023b2v}
\bibinfo{author}{L.~Rauch}, \bibinfo{author}{R.~Schwinger}, \bibinfo{author}{M.~Wirth}, \bibinfo{author}{B.~Sick}, \bibinfo{author}{S.~Tomforde}, \bibinfo{author}{C.~Scholz},
\newblock \bibinfo{title}{Active {{Bird2Vec}}: {{Towards End-to-End Bird Sound Monitoring}} with {{Transformers}}},
\newblock \bibinfo{journal}{CoRR}  (\bibinfo{year}{2023}). \URLprefix \url{https://doi.org/10.48550/arXiv.2308.07121}.
%Type = Article
\bibitem[{Rauch et~al.(2024)Rauch, Schwinger, Wirth, Heinrich, Huseljic, Herde, Lange, Kahl, Sick, Tomforde, and Scholz}]{rauch2024birdset}
\bibinfo{author}{L.~Rauch}, \bibinfo{author}{R.~Schwinger}, \bibinfo{author}{M.~Wirth}, \bibinfo{author}{R.~Heinrich}, \bibinfo{author}{D.~Huseljic}, \bibinfo{author}{M.~Herde}, \bibinfo{author}{J.~Lange}, \bibinfo{author}{S.~Kahl}, \bibinfo{author}{B.~Sick}, \bibinfo{author}{S.~Tomforde}, \bibinfo{author}{C.~Scholz},
\newblock \bibinfo{title}{Birdset: {A Large-Scale Dataset for Audio Classification in Avian Bioacoustics}},
\newblock \bibinfo{journal}{CoRR}  (\bibinfo{year}{2024}). \URLprefix \url{https://doi.org/10.48550/arXiv.2403.10380}.
%Type = Inproceedings
\bibitem[{Denton et~al.(2022)Denton, Wisdom, and Hershey}]{denton2021separation}
\bibinfo{author}{T.~Denton}, \bibinfo{author}{S.~Wisdom}, \bibinfo{author}{J.~R. Hershey},
\newblock \bibinfo{title}{Improving {{Bird Classification}} with {{Unsupervised Sound Separation}}},
\newblock in: \bibinfo{booktitle}{{{ICASSP}} 2022 - 2022 {{IEEE International Conference}} on {{Acoustics}}, {{Speech}} and {{Signal Processing}} ({{ICASSP}})}, \bibinfo{publisher}{IEEE}, \bibinfo{year}{2022}, pp. \bibinfo{pages}{636--640}. \URLprefix \url{https://doi.org/10.1109/ICASSP43922.2022.9747202}.
%Type = Incollection
\bibitem[{Höchst et~al.(2022)Höchst, Bellafkir, Lampe, Vogelbacher, Mühling, Schneider, Lindner, Rösner, Schabo, Farwig, and Freisleben}]{hoechst22}
\bibinfo{author}{J.~Höchst}, \bibinfo{author}{H.~Bellafkir}, \bibinfo{author}{P.~Lampe}, \bibinfo{author}{M.~Vogelbacher}, \bibinfo{author}{M.~Mühling}, \bibinfo{author}{D.~Schneider}, \bibinfo{author}{K.~Lindner}, \bibinfo{author}{S.~Rösner}, \bibinfo{author}{D.~G. Schabo}, \bibinfo{author}{N.~Farwig}, \bibinfo{author}{B.~Freisleben},
\newblock \bibinfo{title}{Bird@{Edge}: {Bird} {Species} {Recognition} at the {Edge}},
\newblock in: \bibinfo{booktitle}{Networked {Systems}}, volume \bibinfo{volume}{13464}, \bibinfo{address}{Cham}, \bibinfo{year}{2022}, pp. \bibinfo{pages}{69--86}. \URLprefix \url{https://doi.org/10.1007/978-3-031-17436-0_6}.
%Type = Incollection
\bibitem[{Bellafkir et~al.(2023)Bellafkir, Vogelbacher, Schneider, M{\"u}hling, Korfhage, and Freisleben}]{bellafkir2023}
\bibinfo{author}{H.~Bellafkir}, \bibinfo{author}{M.~Vogelbacher}, \bibinfo{author}{D.~Schneider}, \bibinfo{author}{M.~M{\"u}hling}, \bibinfo{author}{N.~Korfhage}, \bibinfo{author}{B.~Freisleben},
\newblock \bibinfo{title}{Edge-{{Based Bird Species Recognition}} via {{Active Learning}}},
\newblock in: \bibinfo{booktitle}{Networked {{Systems}}}, volume \bibinfo{volume}{14067}, \bibinfo{publisher}{{Springer Nature Switzerland}}, \bibinfo{address}{{Cham}}, \bibinfo{year}{2023}, pp. \bibinfo{pages}{17--34}. \DOIprefix\doi{10.1007/978-3-031-37765-5_2}.
%Type = Inproceedings
\bibitem[{Vellinga and Planqu{\'{e}}(2015)}]{vellinga2015}
\bibinfo{author}{W.~Vellinga}, \bibinfo{author}{R.~Planqu{\'{e}}},
\newblock \bibinfo{title}{The xeno-canto collection and its relation to sound recognition and classification},
\newblock \bibinfo{publisher}{CEUR-WS.org}, \bibinfo{year}{2015}. \URLprefix \url{https://xeno-canto.org/}.
%Type = Article
\bibitem[{{Van Merri{\"e}nboer} et~al.(2024){Van Merri{\"e}nboer}, Hamer, Dumoulin, Triantafillou, and Denton}]{vanmerrienbour2024}
\bibinfo{author}{B.~{Van Merri{\"e}nboer}}, \bibinfo{author}{J.~Hamer}, \bibinfo{author}{V.~Dumoulin}, \bibinfo{author}{E.~Triantafillou}, \bibinfo{author}{T.~Denton},
\newblock \bibinfo{title}{Birds, {{Bats}} and beyond: Evaluating generalization in bioacoustic models},
\newblock \bibinfo{journal}{CoRR}  (\bibinfo{year}{2024}).
%Type = Inproceedings
\bibitem[{Liu et~al.(2019)Liu, Jiang, He, Chen, Liu, Gao, and Han}]{liu2019radam}
\bibinfo{author}{L.~Liu}, \bibinfo{author}{H.~Jiang}, \bibinfo{author}{P.~He}, \bibinfo{author}{W.~Chen}, \bibinfo{author}{X.~Liu}, \bibinfo{author}{J.~Gao}, \bibinfo{author}{J.~Han},
\newblock \bibinfo{title}{On the variance of the adaptive learning rate and beyond},
\newblock in: \bibinfo{booktitle}{International Conference on Learning Representations}, \bibinfo{year}{2019}.
%Type = Inproceedings
\bibitem[{Huseljic et~al.(2024)Huseljic, Hahn, Herde, Rauch, and Sick}]{huseljic2024fast}
\bibinfo{author}{D.~Huseljic}, \bibinfo{author}{P.~Hahn}, \bibinfo{author}{M.~Herde}, \bibinfo{author}{L.~Rauch}, \bibinfo{author}{B.~Sick},
\newblock \bibinfo{title}{Fast fishing: Approximating bait for efficient and scalable deep active image classification},
\newblock in: \bibinfo{booktitle}{Machine Learning and Knowledge Discovery in Databases. Research Track}, \bibinfo{publisher}{Springer Nature Switzerland}, \bibinfo{address}{Cham}, \bibinfo{year}{2024}.
%Type = Inproceedings
\bibitem[{Huseljic et~al.(2023)Huseljic, Herde, Hahn, and Sick}]{huseljic2023role}
\bibinfo{author}{D.~Huseljic}, \bibinfo{author}{M.~Herde}, \bibinfo{author}{P.~Hahn}, \bibinfo{author}{B.~Sick},
\newblock \bibinfo{title}{Role of hyperparameters in deep active learning},
\newblock in: \bibinfo{booktitle}{Workshop on Interactive Adaptive Learning @ ECML PKDD}, \bibinfo{year}{2023}, pp. \bibinfo{pages}{19--24}.
%Type = Misc
\bibitem[{Kahl et~al.(2022)Kahl, Wood, Chaon, Peery, and Klinck}]{sierra_nevada_stefan_kahl_2022_7050014}
\bibinfo{author}{S.~Kahl}, \bibinfo{author}{C.~M. Wood}, \bibinfo{author}{P.~Chaon}, \bibinfo{author}{M.~Z. Peery}, \bibinfo{author}{H.~Klinck}, \bibinfo{title}{A collection of fully-annotated soundscape recordings from the western united states}, \bibinfo{year}{2022}. \URLprefix \url{https://doi.org/10.5281/zenodo.7050014}.
%Type = Inproceedings
\bibitem[{Hacohen et~al.(2022)Hacohen, Dekel, and Weinshall}]{hacohen2022typiclust}
\bibinfo{author}{G.~Hacohen}, \bibinfo{author}{A.~Dekel}, \bibinfo{author}{D.~Weinshall},
\newblock \bibinfo{title}{Active learning on a budget: Opposite strategies suit high and low budgets},
\newblock in: \bibinfo{booktitle}{International Conference on Machine Learning}, \bibinfo{year}{2022}.
%Type = Inproceedings
\bibitem[{Ash et~al.(2020)Ash, Zhang, Krishnamurthy, Langford, and Agarwal}]{ash2020BADE}
\bibinfo{author}{J.~T. Ash}, \bibinfo{author}{C.~Zhang}, \bibinfo{author}{A.~Krishnamurthy}, \bibinfo{author}{J.~Langford}, \bibinfo{author}{A.~Agarwal},
\newblock \bibinfo{title}{Deep batch active learning by diverse, uncertain gradient lower bounds},
\newblock in: \bibinfo{booktitle}{International Conference on Learning Representations}, \bibinfo{year}{2020}.
%Type = Article
\bibitem[{Stowell(2021)}]{stowell2021}
\bibinfo{author}{D.~Stowell},
\newblock \bibinfo{title}{Computational bioacoustics with deep learning: A review and roadmap},
\newblock \bibinfo{journal}{CoRR}  (\bibinfo{year}{2021}). \URLprefix \url{https://doi.org/10.48550/arXiv.2112.06725}.
%Type = Article
\bibitem[{Navine et~al.(2024)Navine, Denton, Weldy, and Hart}]{navine2024}
\bibinfo{author}{A.~K. Navine}, \bibinfo{author}{T.~Denton}, \bibinfo{author}{M.~J. Weldy}, \bibinfo{author}{P.~J. Hart},
\newblock \bibinfo{title}{All thresholds barred: Direct estimation of call density in bioacoustic data},
\newblock \bibinfo{journal}{Frontiers in Bird Science} \bibinfo{volume}{3} (\bibinfo{year}{2024}). \DOIprefix\doi{10.3389/fbirs.2024.1380636}.
%Type = Inproceedings
\bibitem[{Rauch et~al.(2023)Rauch, Aßenmacher, Huseljic, Wirth, Bischl, and Sick}]{rauch2023GLAE}
\bibinfo{author}{L.~Rauch}, \bibinfo{author}{M.~Aßenmacher}, \bibinfo{author}{D.~Huseljic}, \bibinfo{author}{M.~Wirth}, \bibinfo{author}{B.~Bischl}, \bibinfo{author}{B.~Sick},
\newblock \bibinfo{title}{Activeglae: A benchmark for deep active learning with transformers},
\newblock in: \bibinfo{booktitle}{Machine Learning and Knowledge Discovery in Databases: Research Track}, \bibinfo{publisher}{Springer Nature Switzerland}, \bibinfo{year}{2023}, p. \bibinfo{pages}{55–74}. \URLprefix \url{https://doi.org/10.1007/978-3-031-43412-9_4}.

\end{thebibliography}

\appendix

\end{document}